\documentclass[a4paper]{jpconf}
\usepackage{graphicx}
\usepackage{amsmath}
\begin{document}
\title{Towards the understanding of radial velocity pulsation in roAp stars}

\author{J C Sousa$^{1,2}$, M S Cunha$^1$}

\address{$^1$ Centro de Astrof\'isica da Universidade do Porto, Portugal}
\address{$^2$ Department of Aplied Mathematics, Universidade do Porto, Portugal}

\ead{jsousa@astro.up.pt, mcunha@astro.up.pt}

\begin{abstract}
High-resolution spectroscopic time series of rapidly oscillating Ap stars show evidence for a co-existence of standing and running waves in their atmospheric layers. With the purpose of understanding these observations we have carried out a theoretical analysis of the pulsations in the outermost layers of these stars, starting from the simplest possible model that still retains all important physical ingredients. In our analysis we considered an isothermal atmosphere in a plane-parallel approximation. Moreover we assumed that in the region considered the magnetic pressure is much larger than the gas pressure and, consequently, that the magnetoacoustic wave has decoupled into its acoustic and magnetic components. Using the analytical solutions for the velocity components appropriate to this model we estimate the velocity component parallel to the line of sight averaged over the visible stellar disk. Fitting the latter to a function of the form Acos($\sigma$t+phase), with $\sigma$ the dimensionless oscillation frequency and t the dimensionless time, we derive the amplitude A and the phase for our model as function of height in the atmosphere. 
\end{abstract}

\section{Introduction}

The rapidly oscillating Ap (roAp) stars are cool chemical peculiar stars which are located in the main sequence part of the classical instability strip. They pulsate with periods that vary typically from 5 to 21 minutes, e.g. [1], and have strong large scale magnetic fields, with typical intensities of a few kG. Moreover they have masses of M$=$2M$_\odot$ [2]. Due to these characteristics, the roAp stars offer the opportunity to observe the interaction of p-modes with strong magnetic fields as can be done for no other star but the sun. The scientific interest of these stars has been recognised since the discover of their oscillations and, in the past few years, a several number of observational results have been published. High-resolution spectroscopic studies done for these stars show evidence for a co-existence of standing and running waves in their atmospheric layers, e.g. [1], [3], [4], [5]. 
%\vspace{1cm}
\section{Theoretical Modeling}
%\vspace{1cm}
With the purpose of understanding these observations we have carried out a theoretical analysis of the pulsations in the outermost layers of these stars, starting from the simplest possible model that still retains all important physical ingredients. In our analysis we considered an isothermal atmosphere in a plane-parallel approximation. The photosphere of a roAp star can be located either in the magnetoacoustic, or in the magnetically dominated region, depending on the strength of the magnetic field, and, due to that, depending on the region that is considered, the oscillations in the atmosphere of these stars might look significantly different [6]. In this study we assumed that the region considered is magnetically dominated, i.e., the magnetic pressure is much larger than the gas pressure and, consequently, that the magnetoacoustic wave has decoupled into its acoustic and magnetic components. The magnetic field is considered to have a dipolar geometry and, hence, to vary in intensity and inclination with latitude. All the results presented are for a model of mass M$=$2.0 M$_\odot$ and radius R=2.0 R$_\odot$, M$_\odot$ and R$_\odot$ being, respectively, the mass and the radius known for the sun. The temperature of the isothermal atmosphere is T=8440 K and the atmosphere is matched continuously onto the polytropic interior at a pressure P=7563 gcm$^{-1}$s$^{-2}$, a value motivated by a stellar model with a similar mass and radius. The cyclic oscillation frequency $\nu$ is assumed to be 3.08 mHz, which is above the critical frequency in this atmosphere, and the characteristic magnetic field is assumed to have a strength of 3000 G. The observer is considered to be pole-on in relation to the magnetic field axis. In Fig.1 we show a schematic representation of the star, where $\stackrel{\rightarrow}{B}$ is the local magnetic field, $\alpha$ is the angle between the local magnetic field and the direction of the observer, $\theta$ is the co-latitude and $\phi$ the longitude.

\begin{figure}[h]
\includegraphics[width=15pc]{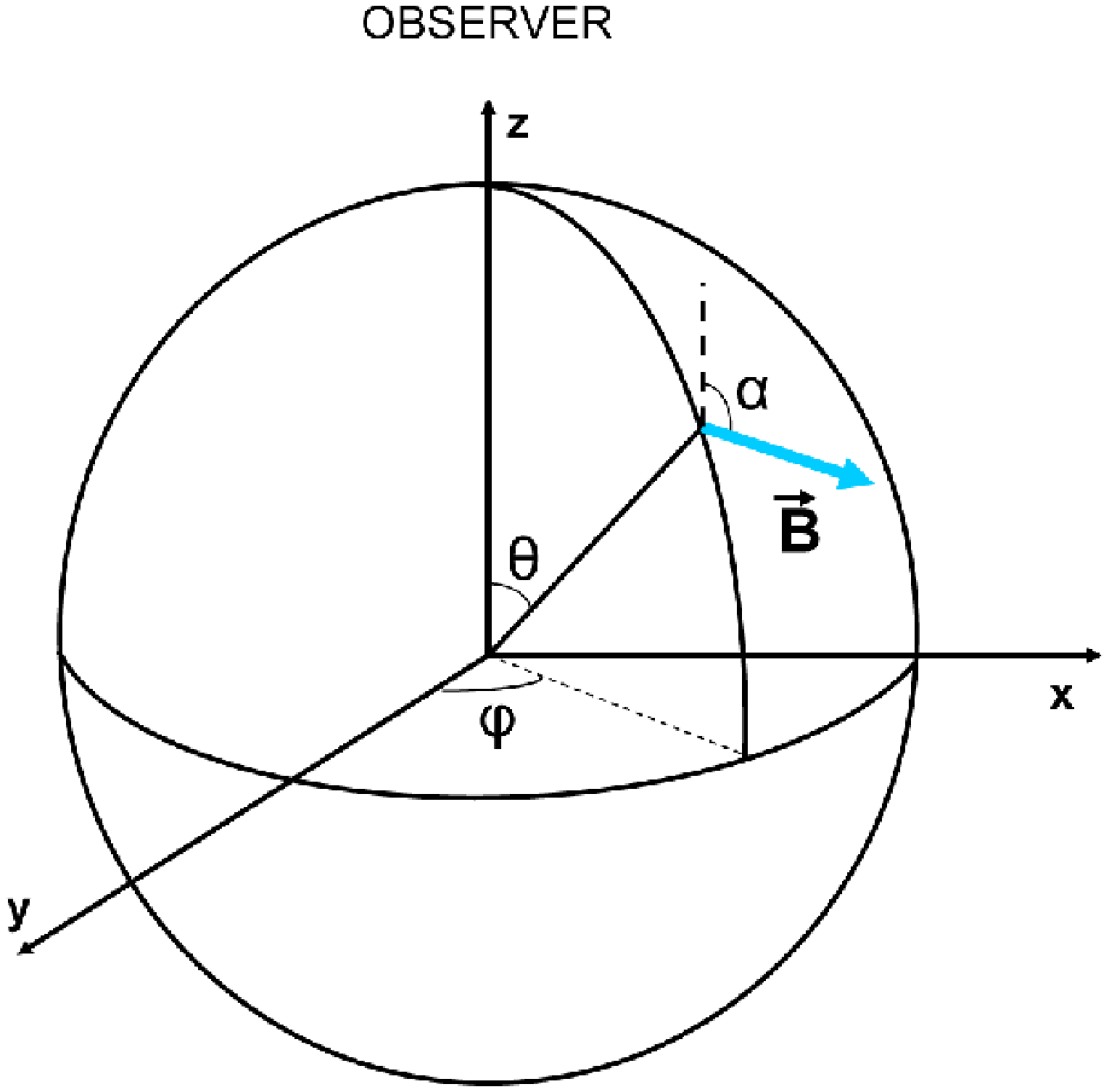}\hspace{4pc}
\begin{minipage}{19pc}
{\label{label}Fig.1 - Schematic representation of the star.$\stackrel{\rightarrow}{B}$ is the local magnetic field, $\alpha$ is the angle between the local magnetic field and the direction of the observer, $\theta$ is the co-latitude and $\phi$ the longitude.}
\end{minipage}
\end{figure}
%\caption{\label{label}Fig.1 - Schematic representation of the star.} 

\vspace{0.5cm}
Starting from the equations for the displacement parallel and perpendicular to the direction of the magnetic field in the region where the magnetic pressure dominates over the gas pressure [7] and using the analytical solutions for the velocity components appropriate to this model we determine the expression for the velocity component parallel to the line of sight averaged over the visible stellar disk, $v_{int}$,

\begin{eqnarray}
v_{int}=\frac{\pi\sigma}{(\pi-\frac{a}{3})}\int_0^{\pi/2}\big[-\frac{A_c(\theta)}{p^{1/2}}\sin(\sigma t+k\eta+\phi_c(\theta))\cos\alpha \nonumber
\end{eqnarray}

\begin{eqnarray}
-\frac{A_m(\theta)}{{\big(p+\frac{b^2}{\gamma C_s\sin(\alpha-\theta)} \big)}^{1/2}}(1-\frac{\sigma^2H^2\rho}{b^2})\sin(\sigma t+\phi_m(\theta))\sin\alpha\big](1-a(1-\cos\theta))d\theta, & &\nonumber
\end{eqnarray}

\vspace{1.5cm}
where
%\vspace{1cm}

\begin{equation}
k={\big[\frac{\sigma^2\rho}{\gamma C_s\cos^2(\alpha-\theta)}-\frac{1}{4H^2} \big]}^{1/2}, \nonumber
\end{equation}

\vspace{1cm}
and $A_c(\theta)$ is the amplitude of the acoustic wave, $\sigma$ is the dimensionless angular frequency, $a$ the limb darkening coefficient, $p$ the dimenssionless pressure, $\phi_c(\theta)$ the phase of the acoustic wave, $\eta$ the dimenssinless depth, $A_m(\theta)$ the amplitude of the magnetic wave, $\gamma$ the first adiabatic exponent, $b$ the magnetic field modulus, $H$ the dimensinless scale height, $\rho$ the dimensionless density, $\phi_m(\theta)$ the phase of the magnetic wave and $C_s$ a constante that is related to the sound speed in the atmosphere.

Fitting $v_{int}$ to a function of the form Acos($\sigma$t+$\phi$), we derive the phase $\phi$,

\vspace{1cm}
\begin{equation}
\phi=\frac{\int_0^{\pi/2} \big( C_c\cos(k\eta+\phi_c(\theta))+C_m\cos(\phi_m(\theta))\big)d\theta}{\int_0^{\pi/2}\big( -C_c\sin(k\eta+\phi_c(\theta))-C_m\sin(\phi_m(\theta))\big)d\theta},\nonumber
\end{equation}

\vspace{1cm} 
and the amplitude A,

\vspace{1cm}
\begin{equation}
{\rm A}=\frac{1}{\sin\phi}\int_0^{\pi/2} \frac{\pi\sigma}{(\pi-\frac{a}{3})}\big[C_c\cos(k\eta+\phi_c(\theta))+C_m\sin(\phi_m(\theta))\big]d\theta\nonumber,
\end{equation}
\vspace{1cm}

for our model as function of height in the atmosphere, where $C_c$ is defined as
\vspace{1cm}
\begin{equation}
C_c=\frac{A_c(\theta)}{p^{1/2}}\cos\alpha(1-a(1-\cos\theta))\sin(2\theta)\nonumber,
\end{equation}
\vspace{1cm}
and
%\vspace{1cm}
\begin{equation}
C_m=\frac{A_m(\theta)}{{\big(p+\frac{b^2}{\gamma C_s\sin(\alpha-\theta)}\big)}^{1/2}}\big(1-\frac{\sigma^2H^2\rho}{b^2}\big)\sin\alpha(1-a(1-\cos\theta))\sin(2\theta)\nonumber.
\end{equation}
\vspace{1cm}
\section{Results}

In Fig. 2 we show the amplitude and phase diagrams as function of height in the atmosphere. The results presented assumed an $A_c(\theta)$ and an $A_m(\theta)$ varying according to,

\begin{equation}
A_c(\theta)=0.1\cos\theta\nonumber,
\end{equation}

\begin{equation}
A_m(\theta)=\sin\theta\nonumber,
\end{equation}
These functional form for the amplitudes were motivated by the results Sousa and Cunha [7]. 

\begin{figure}[!h]
\begin{minipage}{18pc}
\includegraphics[width=18pc]{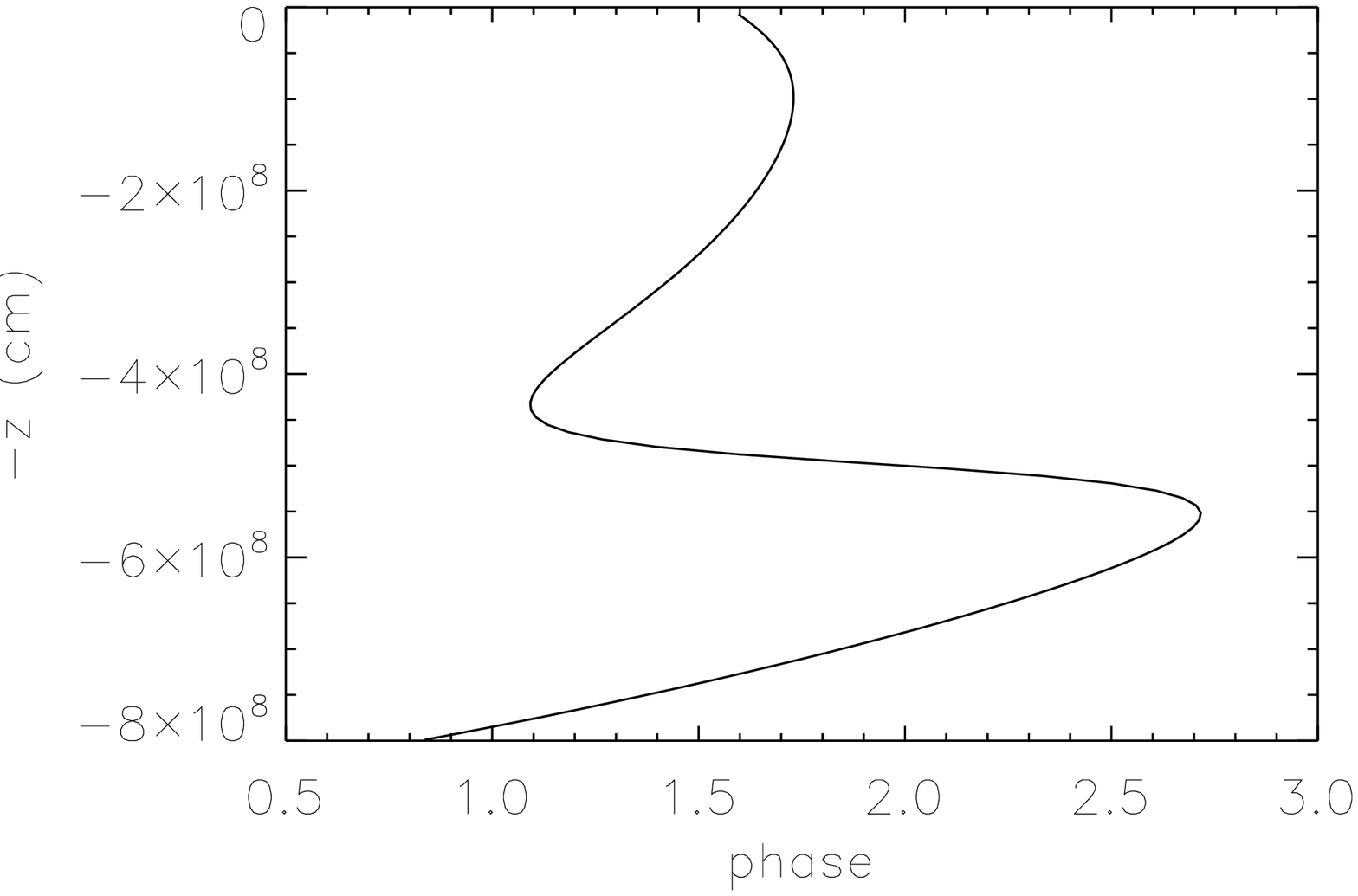}
%\caption{\label{label}Figure caption for first of two sided figures.}
\end{minipage}\hspace{2pc}%
\begin{minipage}{18pc}
\includegraphics[width=18pc]{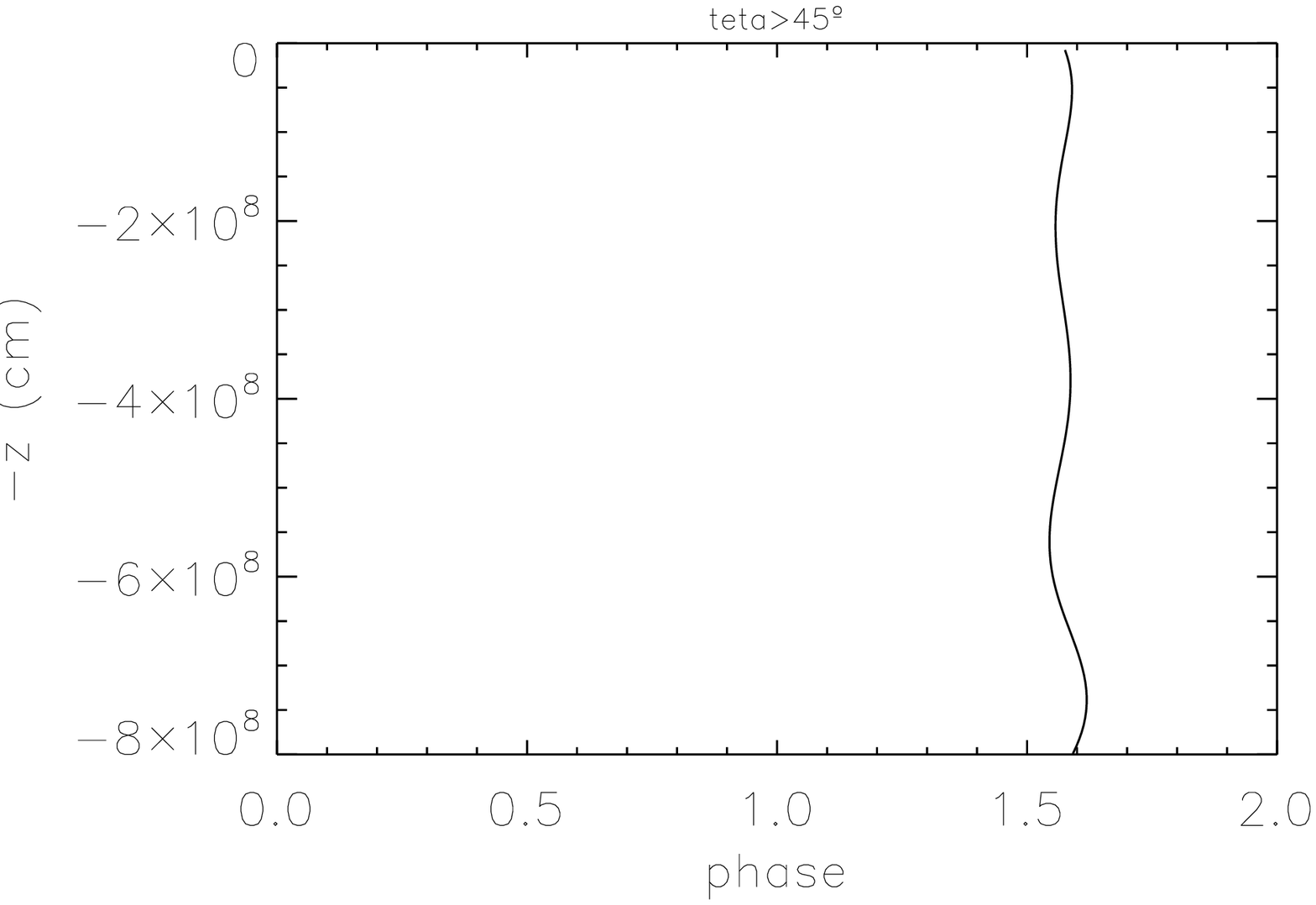}
%\caption{\label{label}Figure caption for second of two sided figures.}
\end{minipage} 
\end{figure}

\begin{figure}[!h]
\begin{minipage}{18pc}
\includegraphics[width=18pc]{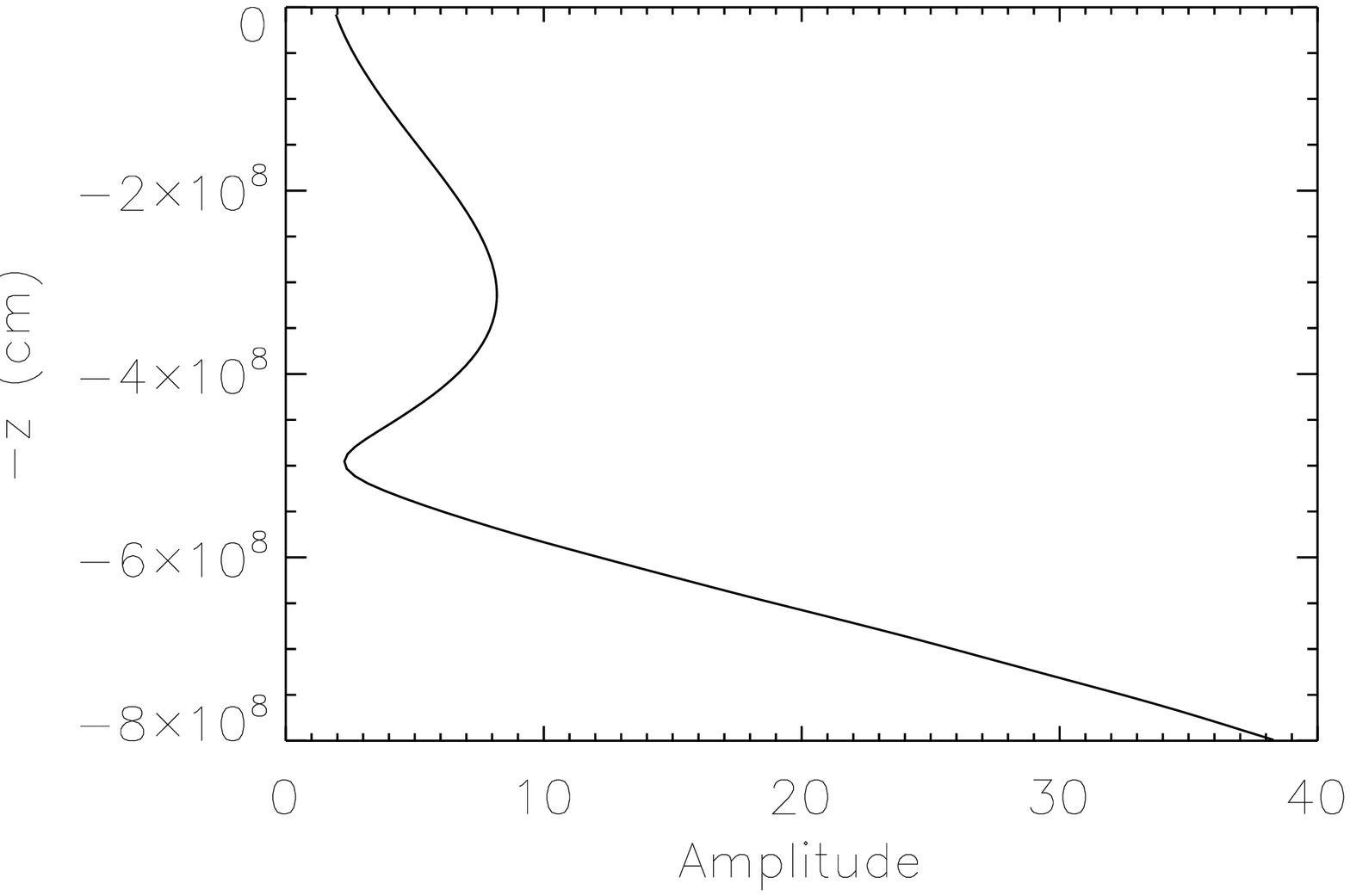}
%\caption{\label{label}Figure caption for first of two sided figures.}
\end{minipage}\hspace{2pc}%
\begin{minipage}{18pc}
\includegraphics[width=18pc]{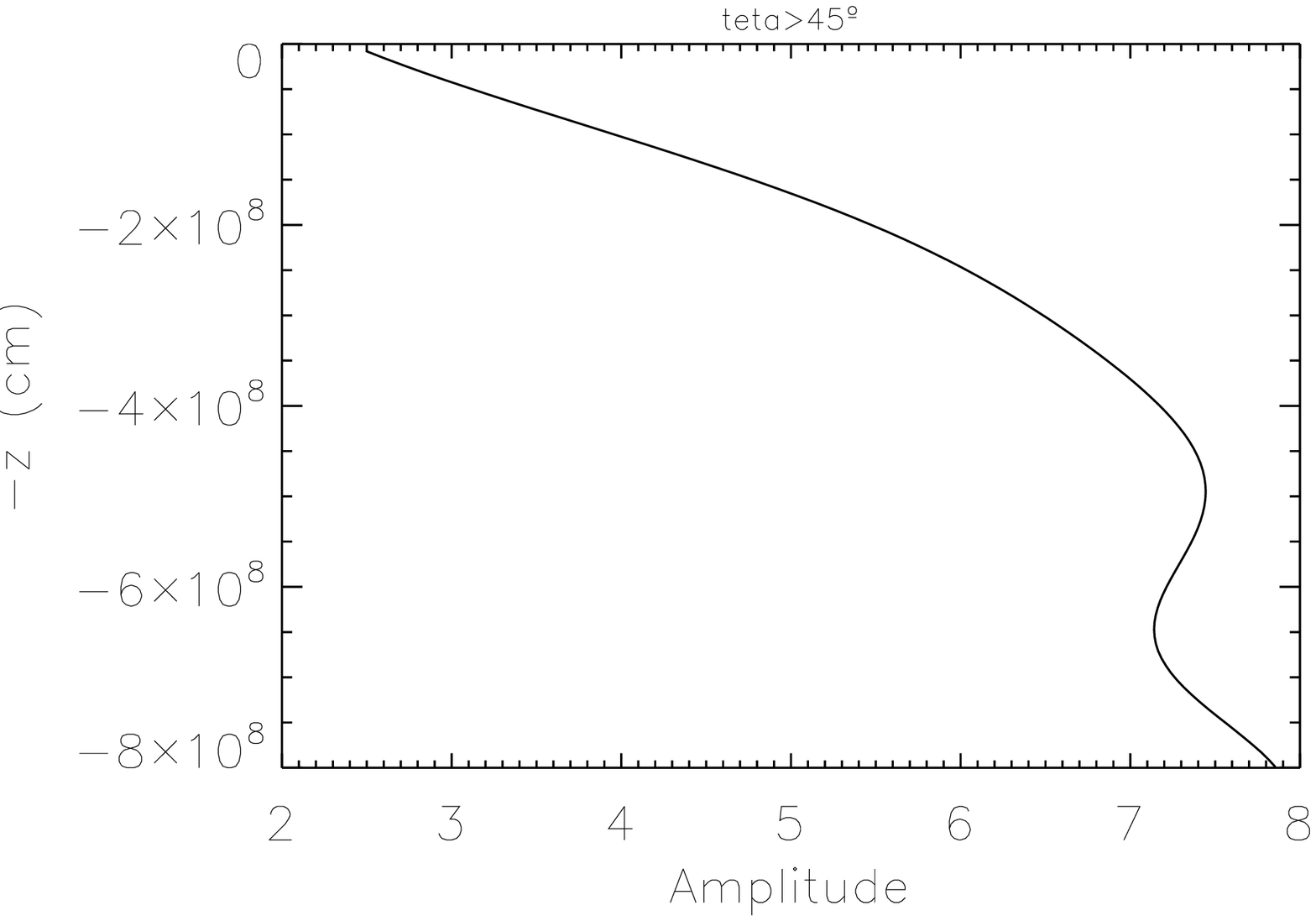}
%\caption{\label{label}Figure caption for second of two sided figures.}
\end{minipage} 
\end{figure}
{\label{label}Fig.2 - Amplitude and phase diagrams as function of heigth in atmosphere. z is the vertical local coordenate, defined to be zero at the photosphere and increase outwardly.In the left panel we show the amplitude and phase derived when the velocity is integrated over the whole stellar disk, $0<\theta<\pi/2$, and in the right panel the amplitude and phase derived when the velocity is integrated over $\pi/4 <\theta<\pi/2$.}

\vspace{1cm}

In the left panel we show the amplitude and phase derived when the velocity is integrated over the whole stellar disk, $0<\theta<\pi/2$, and in the right panel the amplitude and phase derived when the velocity is integrated over $\pi/4 <\theta<\pi/2$. The second case is considered due to the fact that the observed spectral lines, from which the velocity is determined, might originate in particular regions of the stellar surface, rather than being homogenously distributed over the whole surface. It can be seen that in both cases the amplitude shows an overall increase in the outward direction. This is the result of the decreasing density with height. At first sight, the phases show a pattern that seems to indicate the existence of an outwardly running wave, when the phase is decreasing with height, an inwardly running wave, when the phase is increasing with height, and a standing wave when the phase is almost constant. But, in fact, all of these patterns are the result of summing two waves, one acoustic outwardly running wave and one magnetic standing wave, in different proportions. The amplitudes and phases derived from $v_int$, have contributions from the acoustic and magnetic components and depending on which is dominant different results are found. Thus, when the acoustic contribution is dominant we see a running wave propagating outwardly, when the magnetic contribution is dominant we see a standing wave, and when the two waves have similar contributions, the phase is such that it might look like a wave propagating inwardly. So, with this study we suggest that what is observed in high-resolution spectroscopic time series of roAp stars results from the superposition of different contributions of acoustic and magnetic waves.      

\vspace{1cm}
\section{References}
%%%%%%%%%%%%%%%%%%%%%%%%%%%%%%%%%%%%%%%%%%%

\end{document}